# The influence of nitrogen doping and annealing on the silicon vacancy in 4H-SiC

Samuel G. Carter[1,*], Infiter Tathfif [1,2], Charity Burgess[3], Brenda VanMil [3], Suryakanti Debata[4], and Pratibha Dev[4]

[1] *Laboratory for Physical Sciences, College Park, MD 20740, USA*

[2] *Department of Electrical and Computer Engineering, University of Maryland, College Park, MD 20742, USA*

[3] *DEVCOM Army Research Laboratory, Adelphi, MD 20783, USA*

[4] *Department of Physics and Astronomy, Howard University, Washington, D.C. 20059, USA*

**ABSTRACT.** The silicon vacancy ($V_{Si}$) in 4H-SiC at its cubic site (V2-center) has shown significant promise for quantum technologies, due to coherent spin states, the mature material system, and stable optical emission. In these SiC-based applications, doping plays a crucial role. It can be used to control the charge state of $V_{Si}$ and formation of different types of defects. Despite its importance, there has been little research on the effects of doping. In this work, we perform a study of the effects of nitrogen doping and annealing on the photoluminescence (PL), optically-detected magnetic resonance (ODMR) contrast, and dephasing times of ensembles of V2 in epilayers of 4H-SiC. The results show an enhancement of PL that depends on the electron irradiation dose for a given electron concentration, supported by theoretical modeling of the charge state of $V_{Si}$ in the presence of nitrogen. Nitrogen substituted for carbon is shown to very efficiently donate one electron to $V_{Si}$. We also observe that the ODMR contrast can be increased from 0.5% in low doped SiC to 1.5% by nitrogen doping of $10^{17}$ to $10^{18}$ cm$^{-3}$ and annealing at 500-600 °C for 1 hour, with only a 20% decrease in PL compared to unannealed. Some of the improvement in contrast is offset by a reduction in $T_2^*$ at these doping levels, but the estimated cw ODMR shot-noise limited sensitivity is still 1.6 times higher than that of undoped, unannealed SiC.

## I. INTRODUCTION

Color centers in SiC have shown significant potential for both quantum sensing [1–10] and quantum networking [11–15], due to long spin coherence times, good optical properties, and mature fabrication and device capabilities. In particular, the V2 silicon vacancy ($V_{Si}$) in 4H-SiC shows strong potential for quantum networks due to nearly lifetime-limited zero phonon lines (ZPL), even within nanostructures [16,17], and strong potential for sensing due to room temperature operation and the ability to largely remove silicon and carbon isotopes with nonzero nuclear spins. Another potential advantage of color centers in SiC over those in diamond is wafer-scale fabrication and the ease of producing electrical devices with n- and p-type doping, with recent demonstrations of color centers in diodes that have improved linewidths and can be tuned [18–20].

---

[*] Corresponding author: sgcarte@lps.umd.edu

Despite the interest in incorporating color centers like $V_{Si}$ in SiC electrical devices, not much is known about the effects of doping on the optical and spin properties of $V_{Si}$ [21,22]. In a number of experiments, $V_{Si}$ centers are located in the intrinsic region of diodes [20], in part to avoid charge traps or nuclear spins associated with dopants. However, there may be advantages in some cases to having a significant level of doping in the region of $V_{Si}$ centers, such as stabilizing the charge state or incorporating emitters into LEDs [23], and it is important to determine the effects on optical and spin properties.

In particular, it would be beneficial if doping could improve the relatively low contrast in optically-detected magnetic resonance (ODMR) for V2, which is currently a disadvantage for quantum sensing. The contrast is typically defined as the difference between the detected photoluminescence (PL) with and without a radio-frequency (RF) source driving a spin transition, divided by the PL with no RF drive. The continuous wave (cw) ODMR contrast for V2 ensembles at room temperature is typically 0.5% or less [24,25], with the spin transitions saturated. For the current frontrunner in defect-based quantum sensing, the nitrogen-vacancy (NV) center in diamond, the contrast is about 2% for ensembles with all four orientations split apart [26]. Since magnetic field sensitivity is inversely proportional to contrast [27], low contrast is a significant limitation for sensing. There have been previous studies that showed improvements in ODMR contrast from annealing and quenching that seem promising [9].

We perform a study of the optical and spin properties of ensembles of $V_{Si}$ centers for a series of 4H-SiC epilayers with different nitrogen doping levels, anneal temperatures, and electron irradiation doses. In Section II, the experimental methods and samples are described. Epilayer carrier concentration varies from $10^{14}$ to $10^{18}$ cm$^{-3}$, anneal temperatures from unannealed to 900 °C, and irradiation doses from $10^{17}$ to $10^{19}$ cm$^{-2}$ at 1 MeV. Section III presents the experimental results. Room temperature PL and cw ODMR signals are measured for a wide range of parameters in Section IIIA, low temperature PL spectra for a set of lowest and highest doped samples are presented in Section IIIB, and Ramsey measurements of $T_2^*$ for a representative subset are given in Section IIIC. The two main experimental observations of this study are that the integrated PL has a nonlinear dependence on irradiation dose that is apparent for higher doping levels, and that the ODMR contrast increases at anneal temperatures of 500-600 °C, with a larger increase for higher doping levels. The relative figures of merit for cw ODMR and Ramsey magnetometry are estimated for the range of doping levels and anneal temperatures in Section IIID, with modest improvements over undoped, unannealed SiC. In Section IV, we perform density functional theory (DFT) modeling of $V_{Si}$ in the presence of a nearby $N_C$ that helps explain the integrated PL data and provides a better physical picture of this system. Section V is the conclusion.

## II. EXPERIMENT AND SAMPLES

The 4H-SiC samples consist of 10.2–10.8 μm epilayers grown by hot wall chemical vapor deposition on *n+* substrates oriented 4° off the *c* axis. The nitrogen doping of the substrates is estimated at ~7×10$^{18}$ cm$^{-3}$ by secondary ion mass spectrometry. This study uses 5 wafers, with nitrogen doping in the epilayers resulting in carrier concentrations of 1.0×10$^{14}$, 1.1×10$^{15}$, 1.1×10$^{16}$, 1.0×10$^{17}$, 1.0×10$^{18}$ cm$^{-3}$, as measured by Hg-probe CV. The wafers are diced into 1 cm$^2$ squares and sent for 1 MeV electron irradiation at doses of 1×10$^{17}$, 3×10$^{17}$, 1×10$^{18}$, 3×10$^{18}$, and 1×10$^{19}$ cm$^{-2}$. The irradiated samples are then diced into 3×3 squares, with each series of squares annealed in a rapid thermal annealer at 400, 500, 600, 700, 800, and 900 °C for 1 hour in vacuum.

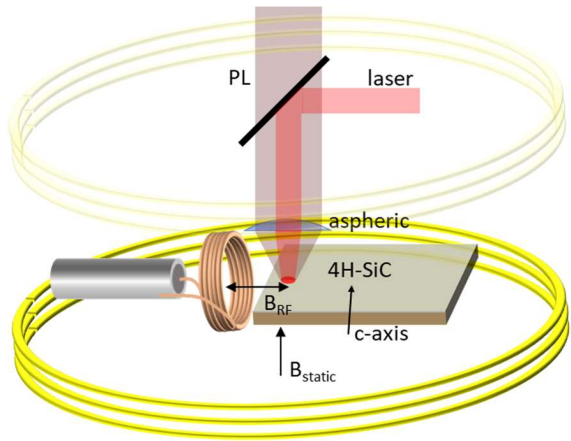

Figure 1. Illustration of the experimental setup for room temperature ODMR, showing the Helmholtz coils (yellow), laser excitation of the SiC sample, and an RF coil near the edge of the sample.

An illustration of the experimental setup for room temperature ODMR is displayed in Fig. 1. Each SiC sample is placed at the center of a pair of home-built Helmholtz coils with a radius of 6.4 cm and about 400 turns each. The magnetic field along the vertical axis is 5.46 mT/A, with experiments performed at either 100 mA (0.55 mT) or 300 mA (1.64 mT). The *c* axis is 4° from the applied magnetic field. Laser excitation was performed with either an 80 MHz femtosecond laser at 770 nm (Coherent Chameleon Discovery) or a cw 785 nm diode laser (Cobolt 06-MLD 785 nm), typically with ~150 mW incident on the sample. No significant difference was observed in the ODMR or PL between these sources. Both lasers can be electrically modulated on/off for pulsed experiments. The laser is reflected off of an 805 nm dichroic long-pass mirror and sent to an aspheric objective lens (NA 0.68) that focuses it onto the sample. A 75 mm focal length lens before the dichroic mirror is used to produce a larger excitation spot of about 20 μm diameter on the sample. PL is collected by the aspheric lens, passes through the dichroic mirror and an 850 nm long pass filter, and then is focused onto a Thorlabs InGaAs detector (either amplified photodiode or avalanche photodiode) with a detection range from 800 to 1700 nm. The wavelength detection window is limited to 850 to 1300 nm by the dichroic mirror and long pass filter.

An RF magnetic driving field is induced by a 5-loop coil of roughly 3 mm diameter shorting the end of a coax cable. This coil is positioned near the edge of the SiC sample and is driven by an SRS SG396 vector signal generator amplified by a Mini-Circuits 10 W amplifier. The RF is modulated either directly through the signal generator or through a Mini-Circuits RF switch. For pulsed experiments the RF switch is used exclusively, and the RF and laser pulses are coordinated with digital outputs of a Swabian Instruments Pulse Streamer. The output of the detector is sent to a lock-in amplifier, with the reference provided by a channel of the Pulse Streamer. The RF is either modulated on/off for cw ODMR and Rabi measurements or modulated with respect to the pulse delay for Ramsey measurements at a frequency of about 5 kHz. Measurements of the integrated PL signal are made by modulating the laser power instead of the RF. ODMR contrast is obtained by dividing the cw ODMR signal by the PL signal.

Low temperature PL measurements are made with samples inside a Montana Instruments Cryostation, with the sample mount at ~4.5 K. The setup outside the Cryostation is similar to the room temperature setup, but a Thorlabs 10X microscope objective with 0.5 NA is used to focus the laser and collect PL. The PL is sent through multi-mode fiber to a spectrometer and either a Princeton Instruments Blaze camera (400 nm – 1050 nm) or Princeton Instruments Nirvana camera (950 nm to 1600 nm).

## III. EXPERIMENTAL RESULTS
### A. Room temperature PL and ODMR

The integrated PL at room temperature can in principle originate from $V_{Si}$, the divacancy, and the nitrogen vacancy centers. Based on low temperature measurements, the PL appears to be primarily from V1 and V2 of $V_{Si}$, particularly for unannealed samples. For $V_{Si}$ only the singly negative charge state is known to emit PL in this wavelength range [28], and room temperature ODMR has only been observed for V2 [3,29]. Thus, the integrated PL will generally be considered a measure of the concentration of $V_{Si}^-$. In Fig. 2, the integrated PL signal is plotted as a function of electron irradiation dose for the doping series, all unannealed. The concentration of $V_{Si}$ in any charge state is expected to be proportional to the dose. The linear dependence on dose for N-doping of $10^{14}$ – $10^{16}$ cm$^{-3}$ indicates that low doped samples have a reasonably high, constant fraction of the monovacancy in the bright charge state of $q = -1$, with electrons provided by sources other than N-doping. For doping levels of $10^{17}$ and $10^{18}$ cm$^{-3}$, the PL is relatively weak for low doses but gets stronger at higher doses, exceeding the low-doped PL at particular values and giving a nonlinear dependence on dose. The data appear to show that PL is weak when the concentration of $V_{Si}$ is low compared to the donor concentration, with $V_{Si}$ likely in the $-2$ state, and maximum at an optimum doping, with $V_{Si}$ in the $-1$ state. This is corroborated by our theoretical calculations that show that the monovacancies adopt a $q = -2$ charge state for high doping levels. The "hump" in the $10^{17}$

cm$^{-3}$ doping series at a dose of about $10^{18}$ cm$^{-2}$ appears to show the optimum ratio of dose to doping for the $-1$ charge state. We note that the $10^{17}$ cm$^{-3}$ doping samples seem to have weaker PL overall, perhaps due to growth differences, although a fair comparison is hard to make. It is also important to note that the substrates are doped roughly an order of magnitude higher than the $10^{18}$ cm$^{-3}$ doping sample, and we do not observe PL from the substrate at these doses, consistent with the idea that only $V_{Si}^{2-}$ is present. More discussion concerning the charge state of $V_{Si}$ in the presence of nitrogen dopants will be provided in the theory section.

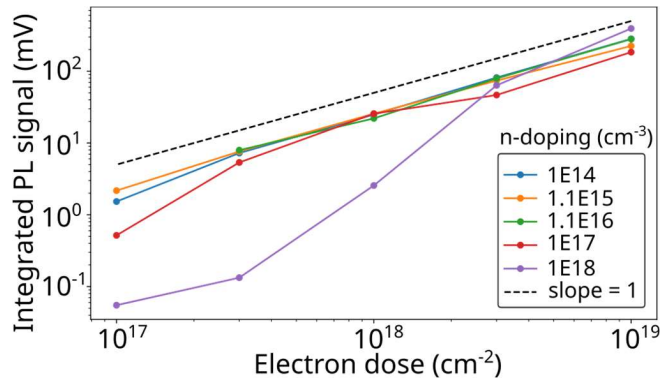

Figure 2. Integrated room temperature PL signal vs. electron irradiation dose for the nitrogen doping series, with no anneal. The excitation laser is 785 nm with 130 mW power. For reference, the dashed line shows a linear relationship between signal and dose.

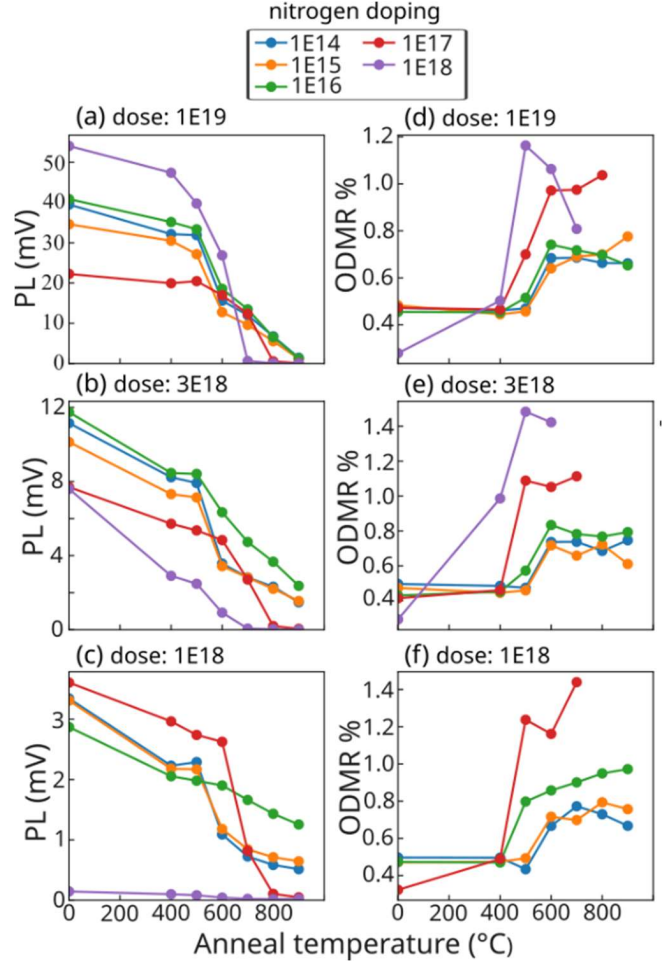

Figure 3. (a-c) Room temperature integrated PL and (d-f) ODMR signals as a function of anneal temperature for different doping levels and for electron irradiation doses of (a,d) $1\times10^{19}$, (b,e) $3\times10^{18}$, and (c,f) $1\times10^{18}$ cm$^{-2}$. Data for unannealed samples are plotted at 0 °C. Data are taken at a static magnetic field of 0.55 mT, with 770 nm, 150 mW laser excitation.

The effects of annealing on integrated room temperature PL are plotted in Fig. 3(a-c) for each doping and for electron irradiation doses from $1\times10^{19}$ to $1\times10^{18}$ cm$^{-2}$. In each sample there is a monotonic decrease in PL with the anneal temperature, typically with the sharpest decrease at about 600 °C. This decrease is likely due to conversion of $V_{Si}$ to $V_C C_{Si}$, observed previously [30,31]. At higher anneal temperatures the decrease continues but more strongly for higher doping and higher doses. For the highest dose, the PL is a few % or less than its unannealed value after a 900 °C anneal for all doping while it is 17-45% for a dose of $1\times10^{18}$ cm$^{-2}$ and doping of $1\times10^{14}$ to $1\times10^{16}$ cm$^{-3}$. Part of this decrease at high doses may be due to $V_C V_{Si}$ formation, which may be more probable for high doses, due to a smaller average distance between $V_C$ and $V_{Si}$. For the higher two doping concentrations, the PL is less than 3% of unannealed values at 800 °C for all doses. We expect that part of the sharp decrease for higher doped samples at these temperatures is due to conversion into $N_C V_{Si}$ [32], which we also observe with low temperature PL measurements.

Representative examples of ODMR spectra at low and high RF power are displayed in Fig. 4(a), in this case for the highest doping, highest electron irradiation, and an anneal temperature of 500 °C. The peak labeled 1 corresponds to the $m_s = +1/2$ to $+3/2$ transition. The other main peak, labeled 2, corresponds to the $m_s = -1/2$ to $-3/2$ transition, which is near the anticrossing as well as other peaks. Peaks 1' and 2' are at half the frequencies of peaks 1 and 2, respectively. They are the result of 2nd harmonics of the drive frequency, coming from the RF source and amplifier, that drive transitions 1 and 2, respectively. We expect peak 3 comes from a 2-photon transition of $m_s = -1/2$ to $+3/2$, studied previously [29], and peak 4 corresponds to the $m_s = +1/2$ to $-3/2$ transition. Peak 1 is studied here, as it is a straightforward $\Delta m_s = 1$ transition far from an anticrossing. At an RF power of 25 dBm, the transition is strongly power broadened, with a maximum ODMR contrast of ~1.1%. At 0 dBm there is little power broadening, and the presence of hyperfine side peaks associated with [29]Si next-nearest-neighbors (NNN) [33] are observed. The linewidths are broader than those observed for lower doped samples, where the hyperfine side peaks can be clearly resolved. This effect is studied in more detail later in this article using Ramsey experiments to measure $T_2^*$.

For ODMR contrast comparisons, it is beneficial to ensure the Rabi frequency stays constant, preventing drift in the contrast. In Fig. 4(b), Rabi oscillations are measured for the same sample as in Fig. 4(a), along with a decaying exponential fit: $A \cdot \exp[-(t/\tau)^n] \cdot \cos(2\pi f t + \phi) + off$. The measurement is performed at a pulse repetition period of 7.5 μs, with a 4 μs laser pulse for initialization and readout, and a variable length RF pulse about 1 μs after the laser pulse. The 30 dBm RF pulse is modulated on and off at a period of 195 μs. The fit gives a Rabi frequency $f$ of 7.76 MHz, with a decay time $\tau$ and decay exponent $n$ of 490 ns and 0.4, respectively. For the ODMR data in Fig. 3(d-f), the Rabi frequency was maintained between 4 and 4.5 MHz, typically with an RF power of 25 dBm. The Rabi frequency is strong enough to drive the hyperfine NNN [29]Si peaks reasonably well, giving nearly the maximum ODMR contrast.

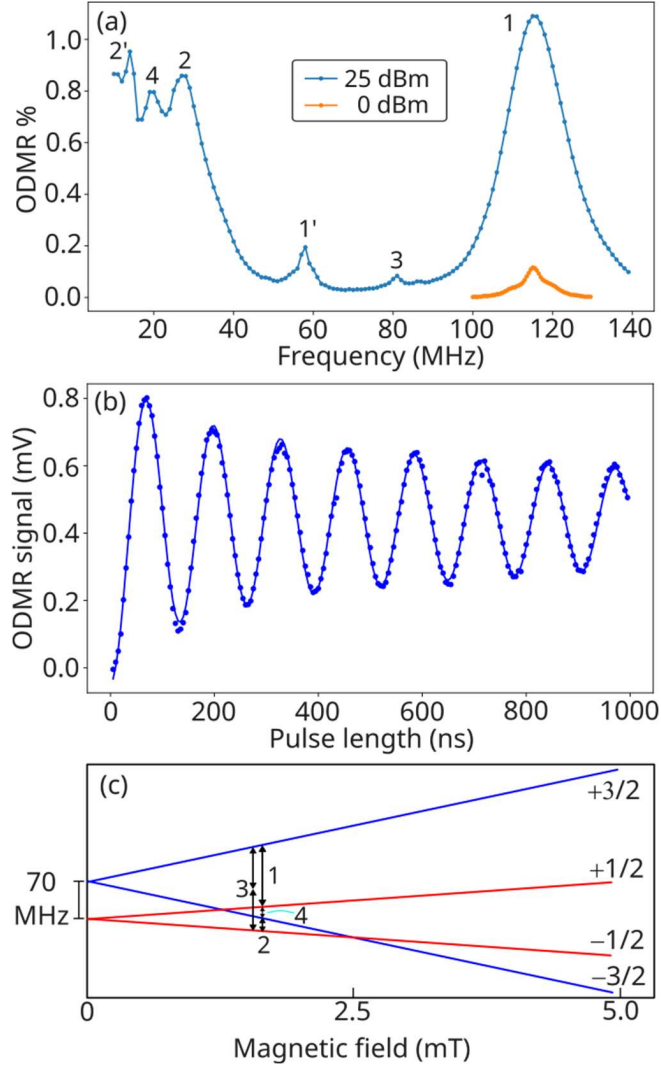

Figure 4. (a) Room temperature cw ODMR spectra at two RF powers, with peaks labeled according to transitions shown in (c). (b) Rabi oscillations for an RF power of 30 dBm. The line is a fit of the data points to a decaying cosine function, $A \cdot \exp[-(t/\tau)^n] \cdot \cos(2\pi f t + \phi) + off$. Data in (a) and (b) is taken at 1.64 mT on a sample doped at $1\times10^{18}$ cm$^{-3}$, irradiated at $1\times10^{19}$ cm$^{-2}$, and annealed at 500 °C. (c) Energies of the $S = \frac{3}{2}$ quartet as a function of magnetic field along the c-axis, with the black arrows indicating transitions observed in (a).

The ODMR contrast in Fig. 3(d) at the highest electron irradiation dose starts at about 0.5% for unannealed samples except for the $10^{18}$ cm$^{-3}$ doped sample, which starts at 0.28%. There is a jump in the contrast at 500-600 °C for all samples, but the contrast increases to higher levels and jumps at a lower temperature for the $10^{17}$ and $10^{18}$ cm$^{-3}$ samples. The contrast reaches 1.16% for the $10^{18}$ cm$^{-3}$ sample at 500 °C, which is about 2.5 times improvement over the contrast of the unannealed, lowest-doped $10^{14}$ cm$^{-3}$ sample. Similar behavior is observed in Fig. 3(e) for an electron irradiation dose of $3 \times 10^{18}$ cm$^{-2}$, but for the $10^{18}$ cm$^{-3}$ doped sample the jump in contrast begins at 400 °C and goes up higher. At an electron irradiation dose of $1 \times 10^{18}$ cm$^{-2}$ in 3(f), the results look similar to those for $1 \times 10^{19}$ cm$^{-2}$, except that the $10^{18}$ cm$^{-3}$ doped sample gives too little PL to measure ODMR, and the $10^{17}$ and $10^{16}$ cm$^{-3}$ samples give results similar to the $10^{18}$ and $10^{17}$ cm$^{-3}$ samples in 3(d). The contrast behavior with anneal temperature seems to depend on the ratio of doping to dose, at least over some range. The increase in contrast with anneal temperature comes at the cost of reduced PL although this reduction is not too high at 500-600 °C where the jump in contrast occurs.

### B. Low temperature PL

At low temperature the ZPL of different defects can be observed, making it clearer which defects are present. Figure 5 plots the PL spectra at ~4 K for the unannealed $10^{14}$ cm$^{-3}$ doped sample and the $10^{18}$ cm$^{-3}$ doped samples for the series of anneal temperatures, all at an electron dose of $10^{19}$ cm$^{-2}$. For this measurement, the laser excitation and PL collection are nearly along the c-axis, which makes emission lines with a dipole moment along the c-axis (i.e. V1 and V2 ZPL) relatively weak [34]. The V1' ZPL at 858 nm, with dipole moment perpendicular to c, is strong while V1 and V2 at 861 nm and 916 nm, respectively, are weak [28]. The phonon sideband emission at longer wavelengths extends out past 1 μm, limited here by the Si CCD. There is not a clear change in the ZPL wavelength or linewidth between the $10^{14}$ cm$^{-3}$ doped sample and the $10^{18}$ cm$^{-3}$ doped samples. As the anneal temperature increases for the highly doped $10^{18}$ cm$^{-3}$ sample, all of the PL in this wavelength range decreases with anneal temperature, consistent with room temperature integrated PL measurements. However, there is an increase in the V2 ZPL peak at the 500 °C anneal, even as the phonon sideband decreases. The fact that the ODMR contrast also jumps up at this anneal temperature may indicate a connection between these behaviors.

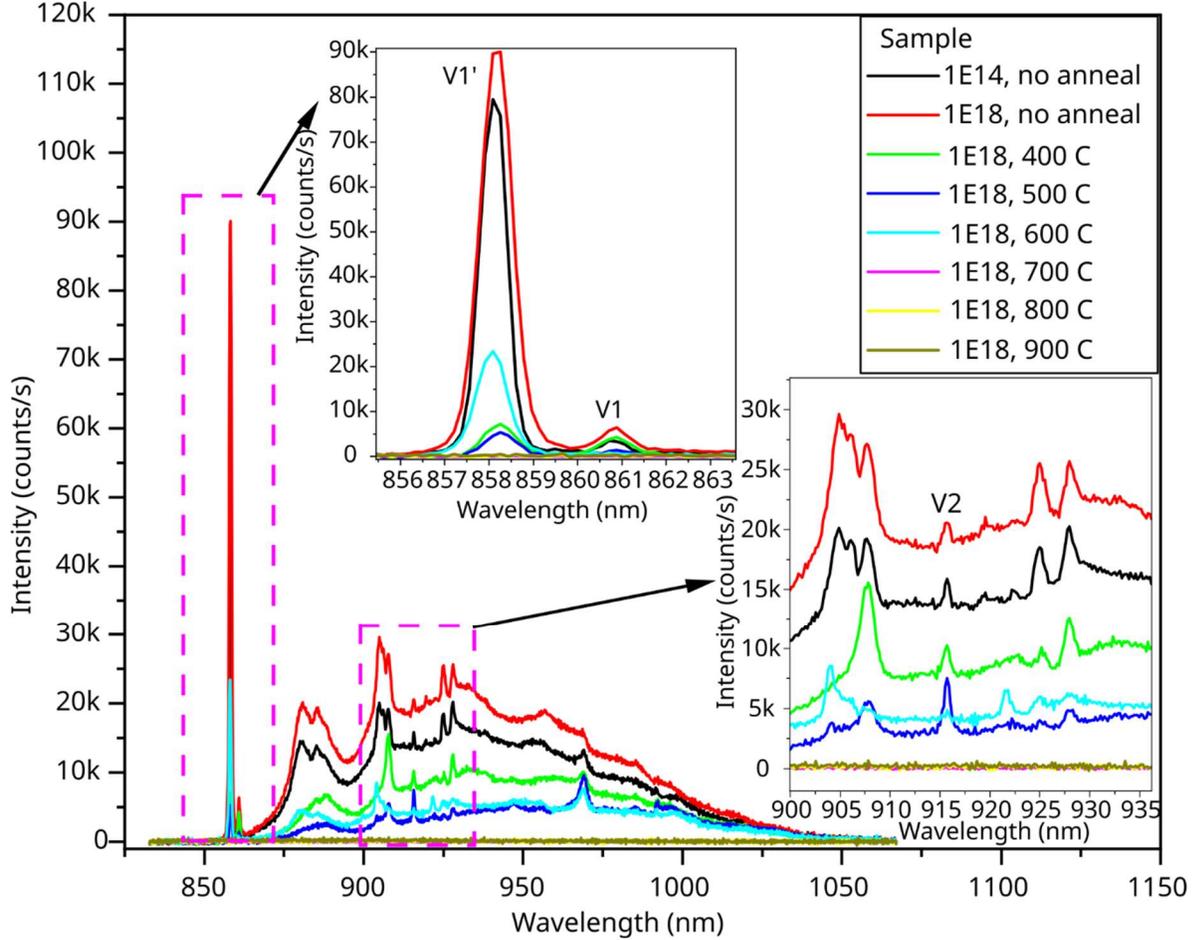

Figure 5. Low temperature PL spectra with a Si CCD and 785 nm excitation for the unannealed samples with nitrogen doping of $10^{14}$ cm$^{-3}$ and $10^{18}$ cm$^{-3}$ for a series of anneal temperatures, all electron irradiated at $10^{19}$ cm$^{-2}$. Insets show expanded views of the V1 and V2 ZPLs.

### C. Dephasing time measurements

The $T_2^*$ dephasing times for a representative subset of the samples were obtained using Ramsey measurements, similar to the Rabi measurements discussed earlier, but with a sequence of two RF pulses separated by a variable delay $\tau$. Figure 6(a) shows the pulse sequence for the measurement and reference. The first pulse induces a $\pi/2$ rotation to the ensemble, creating a superposition between $m_s = +1/2$ and $+3/2$. The 2$^{nd}$ $\pi/2$ pulse results in an overall $\pi$ rotation to the top of the Bloch sphere if the superposition retains the same phase during the delay. As the delay is varied, phase evolution in the frame of the RF drive frequency, including dephasing, is observed. The reference pulse sequence uses a long delay of about 2 µs, coming just before the next initialization/readout pulse, where all coherence is lost. Figure 6(b) plots the Ramsey measurements for N-doping of $10^{14}$ and $10^{18}$ cm$^{-3}$, both with the highest electron dose and a 600

°C anneal. The RF π/2 pulses are 35 and 37 ns for the samples, respectively, corresponding to Rabi frequencies of about 7 MHz. The $10^{14}$ cm$^{-3}$ doped sample shows clear oscillations at 4.8 MHz, corresponding to the $^{29}$Si hyperfine NNN peaks, and an exponential decay time of 213 ns, similar to previous ensemble measurements in natural abundance SiC [24], in which dephasing times are limited by the bath of $^{13}$C and $^{29}$Si nuclear spins. The $10^{18}$ cm$^{-3}$ doped sample shows heavily damped oscillations due to a faster decay time of 58 ns, presumably due to interaction with electron or nuclear spins from the high density of nitrogen atoms.

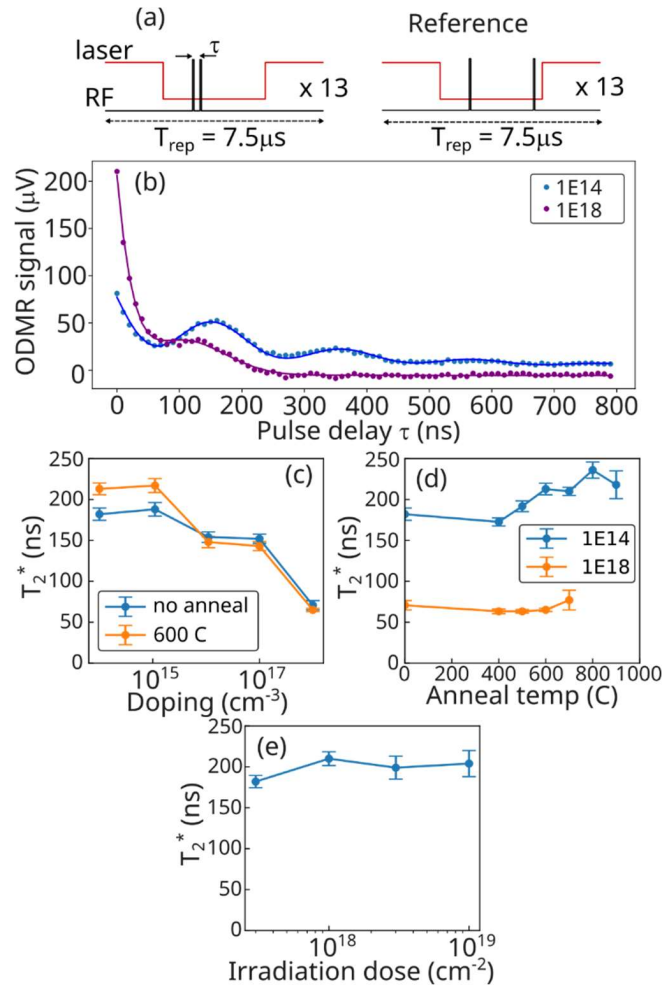

Figure 6. (a) Pulse sequence for Ramsey measurements, showing the measurement sequence and reference sequence. (b) Ramsey measurement for the $10^{14}$ and $10^{18}$ cm$^{-3}$ nitrogen doped samples, irradiated at $10^{19}$ cm$^{-2}$ and annealed at 600 °C. Solid circles represent the data points, and lines represent the fits. (c) Ramsey $T_2^*$ vs. doping for an electron dose of $10^{19}$ cm$^{-2}$ with and without a 600 °C anneal. (d) Ramsey $T_2^*$ vs. anneal temperatures at the lowest and highest doping for an electron dose of $10^{19}$ cm$^{-2}$. (e) Ramsey $T_2^*$ vs. irradiation dose for $10^{14}$ cm$^{-3}$ doping and no anneal.

The trend toward shorter dephasing times with higher doping is plotted in Fig. 6(c), showing fairly modest decreases in $T_2^*$ until the highest doping. The dephasing time is likely dominated by the

$^{13}$C and $^{29}$Si nuclear spin bath until the density of N is high enough to have an effect. We expect the influence of N-doping on $T_2^*$ would be clearer at lower doping levels for isotopically purified SiC. There is also a small increase in $T_2^*$ for a 600 °C anneal for the two lowest doping concentrations. This is measured for a series of anneals in Fig. 6(d), showing a modest increase of ~25% in the dephasing time at the highest anneal temperatures for the $10^{14}$ cm$^{-3}$ doping and no clear effect for the $10^{18}$ cm$^{-3}$ doping. One might also expect higher irradiation doses to result in lower values of $T_2^*$, due to defect-defect spin interactions, as has been observed for $T_2$ [35]. Figure 6(e) plots $T_2^*$ vs irradiation dose for $10^{14}$ cm$^{-3}$ doping and no anneal with no clear effect, indicating that hyperfine with the $^{13}$C and $^{29}$Si nuclear spin bath still dominates even at the highest doses.

### D. Sensing analysis

To quantify the effects of annealing and doping on magnetic field sensing, we consider expressions for shot-noise-limited sensitivity for both cw ODMR and Ramsey measurements, based on Ref. [27]. For cw ODMR, the sensitivity is calculated at the point of maximum slope, which is typically written in terms of a Lorentzian linewidth. With the presence of poorly-resolved hyperfine side peaks, the line shape is more complex, and we formulate the sensitivity in terms of the maximum slope of the ODMR contrast, $\alpha = \max(dC/d\nu)$, where $C$ is the ODMR contrast and $\nu$ is the RF frequency. This gives a simple expression,

$$\eta_{cw} = \frac{\hbar}{g_e \mu_B} \frac{1}{\alpha \sqrt{R}}, \quad (1)$$

in which $g_e = 2$ is the electron g-factor, $\mu_B$ is the Bohr magneton, and $R$ is the photon detection rate. For Ramsey measurements, with initialization/readout times much longer than $T_2^*$, the sensitivity is given by

$$\eta_{Ramsey} = \frac{\hbar}{g_e \mu_B} \frac{1}{C e^{-1} T_2^* \sqrt{R}}. \quad (2)$$

Here, we are interested in the expected changes in sensitivity compared to undoped, unannealed samples commonly used, so we examine relative figures of merit for cw ODMR and Ramsey measurements. For cw ODMR, the relative figure of merit is defined as $F_{cw} = \alpha \sqrt{PL}/[\alpha^{(0)} \sqrt{PL^{(0)}}]$, where the superscript (0) indicates values for the undoped, unannealed sample. Similarly, for Ramsey sensing, the relative figure of merit is $F_{Ramsey} = CT_2^*\sqrt{PL}/C^{(0)}T_2^{*(0)}\sqrt{PL^{(0)}}$.

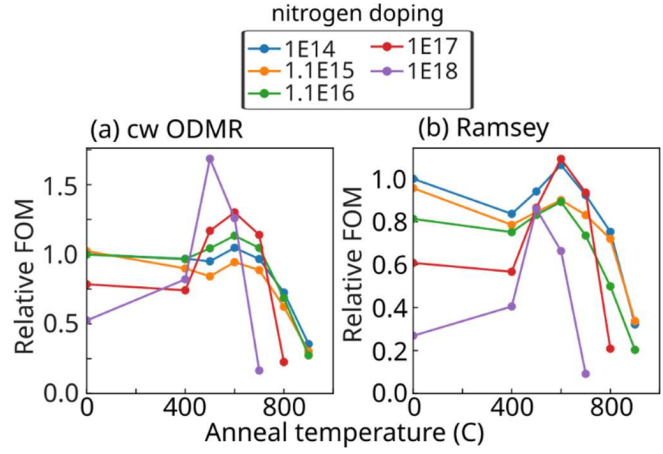

Figure 7. Relative figures of merit for sensing with (a) cw ODMR and (b) Ramsey measurements for the $10^{19}$ cm$^{-2}$ electron irradiation. The figures of merit are relative to the unannealed, lowest doped sample.

Figure 7 plots (a) $F_{cw}$ and (b) $F_{Ramsey}$ for the highest irradiation dose as a function of anneal temperature for the doping series. Only the highest dose is considered as the increased PL should give better sensitivity, with lower doses giving only marginal increases in contrast and no improvements in $T_2^*$. For cw ODMR, the point of maximum slope is obtained for each data point by taking ODMR for a series of powers and differentiating with respect to frequency. For Ramsey measurements, the value of $T_2^*$ was not measured for every point. Instead, $T_2^*$ was estimated for some conditions, based on the trends in Fig. 6(c-d). For the doping concentrations of $10^{14}$ cm$^{-3}$ and $10^{18}$ cm$^{-3}$, $T_2^*$ is measured for each anneal. For the doping of $10^{15}$ cm$^{-3}$, $T_2^*$ is estimated to be the same as those of $10^{14}$ cm$^{-3}$ for missing anneal temperatures since there is no significant difference between these doping concentrations in Fig. 6(c). For the concentrations of $10^{16}$ cm$^{-3}$ and $10^{17}$ cm$^{-3}$, $T_2^*$ is estimated as the average of no anneal and 600 °C for the missing anneal temperatures since there appears to be no significant difference with annealing at these doping concentrations. We also use the cw PL and contrast values for $F_{Ramsey}$, which will in reality be different, but their ratios with the undoped, unannealed values should still be accurate.

The cw sensing figure of merit shows modest peaks at 600 °C for $10^{14}$ – $10^{17}$ cm$^{-3}$, with the largest for $10^{17}$ cm$^{-3}$. For $10^{18}$ cm$^{-3}$, the peak occurs at 500 °C, with a peak value of 1.7. The peaks occur as the contrast jumps up at these temperatures and before the PL decreases too much at higher temperatures. The Ramsey figure of merit $F_{Ramsey}$ also shows peaks at 500 °C and 600 °C, but the peaks are either barely above 1 or below 1. The lack of a significant improvement in $F_{Ramsey}$ for doped samples is the result of the reduction in $T_2^*$. For cw ODMR sensing, the corresponding increase in linewidth has less impact, since the RF drive significantly broadens the ODMR linewidth at powers that give the maximum slope.

## IV. FIRST PRINCIPLES CALCULATIONS

### A. Methods

Spin-polarized density functional theory (DFT)-based calculations were performed using the Quantum-ESPRESSO package [36]. We used the generalized gradient approximation (GGA) [37] of Perdew-Burke-Ernzerhof (PBE) [38]. In this work, we report the results for the V2-center in a $6 \times 6 \times 2$ (576-atoms) supercell of 4H-SiC. The Brillouin-zone was sampled with a Γ-centered, $2 \times 2 \times 2$ k-points grid according to Monkhorst-Pack method [39]. Since nitrogen can n-dope the crystal by either substituting a C-atom or a Si-atom, we first estimated the defect formation energies for simple nitrogen-substitutional defects − $N_C$ and $N_{Si}$ − in a 576-atoms supercell using the following formula:

$$E_{form}(Defect = X) = E_{Total}(X) - [E_{Total}(ideal) - \mu_{Si\,or\,C} + \mu_N] \qquad (3)$$

where the substitutional defect, $X$, is either $N_C$ or $N_{Si}$. $E_{Total}(ideal)$ is the total energy of the ideal 576-atoms supercell, $E_{Total}(X)$ is the total energy of a crystal with a neutral substitutional defect and the $\mu_i$'s (i = Si, C, N) are the chemical potentials for different elements. For carbon and silicon-rich conditions, we used bulk graphite and silicon crystal (fcc), respectively, as references. $N_2$ is used to obtain the reference energy for nitrogen.

TABLE I. Defect-formation energies of simple nitrogen substitutionals in carbon and silicon-rich conditions.

| Defect | $E_{form}$ (eV) C-rich/Si-poor | $E_{form}$ (eV) Si-rich/C-poor |
|---|---|---|
| $N_C$ | -1.59 | -1.91 |
| $N_{Si}$ | 4.44 | 4.77 |

## B. Results

In both carbon- and silicon-rich conditions, we find that the neutral $N_C$ defect has a much lower formation energy than the neutral $N_{Si}$ defect (see Table I). Figure 8(a) is a plot of the density of states (DOS) for the $N_C$−doped 4H-SiC supercell around the Fermi level, showing that $N_C$ is indeed an n-dopant and that it does not introduce localized (sharp) states in the bandgap. The charge-density plots in Fig. 8(b) show the filled states that are just below the conduction band edge. These states are mostly bulk-like in nature, which explains why $N_C$ stabilizes the negative charge state of $V_{Si}$. Hence, we used $N_C$ in the remainder of the calculations to n-dope the defective 4H-SiC crystal. The silicon monovacancy itself was created at a cubic site in the supercell center.

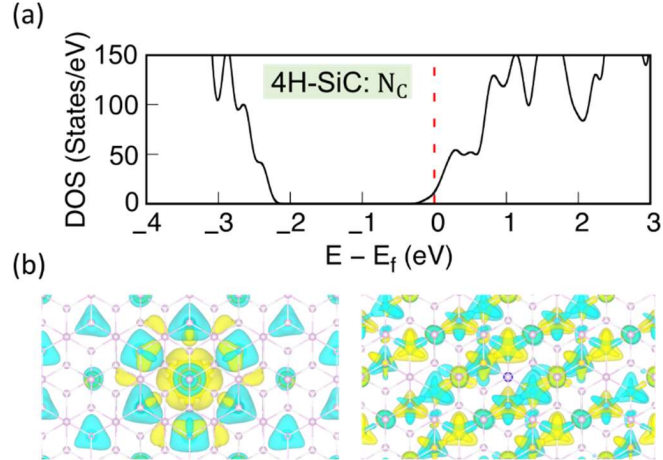

Figure 8. (a) Density of state (DOS) plot around the Fermi level shows the n-doping of 4H-SiC in the presence of $N_C$. (b) The charge density isosurface plots for the two filled states (*at the Γ-point*) just below conduction band edge show the delocalized character of these filled states. The state on the left is a non-degenerate filled state, while that on the right is one of the doubly degenerate states below Fermi level. Pink, white and blue spheres are silicon, carbon and nitrogen atoms. Yellow (teal) colors are used for positive (negative) isovalues of the plotted quantity, $[sign[\psi] \times |\psi|^2]$, where $\psi$ is the selected state.

An $N_C$ substitutional can be created at a number of different sites in the supercell relative to the position of the silicon monovacancy. We chose only those sites that go beyond the first shell of four carbons around $V_{Si}$ to avoid creating an NV-center in the crystal, which is distinct from the $V_{Si}$ defect that is of central interest to this work. In this proof-of-principle exploration of the effects of nitrogen doping on the properties of $V_{Si}$, the following four inequivalent sites [see Fig. 9(a)] were chosen for nitrogen substitution: (i) site $S1$: nearest-neighboring carbon to the first shell of the carbon-atoms, vertically below $V_{Si}$ (on the $C_3$-axis), (ii) site $S1'$: next-nearest carbon to the first shell of carbons, (iii) Site $S2$: farthest possible carbon along the diagonal, and (iv) Site $S2'$: farthest possible carbon on the $C_3$-axis. Overall, our results for the two sites closer to $V_{Si}$ ($S1$ and $S1'$) are qualitatively the same, while they are distinct from the properties of $V_{Si}$ when the nitrogen dopant is farther away (i.e., at the $S2$ and $S2'$ sites). In the presence of $V_{Si}$, the formation energy of $N_C$ at the $S1$ site is lower than $N_C$ at the $S1'$, $S2$ and $S2'$ sites by 62.8 meV, 190.2 meV, and 208.1 meV, respectively. This means that $N_C$ may preferentially form at the $S1$ site and/or migrate to this site upon annealing. For all four sites, we find that $N_C$ readily donates its electron to $V_{Si}$. The monovacancy becomes a negatively charged, spin-3/2 defect, with three more electrons in one spin channel (majority spin ↑) than the other (minority spin ↓). The defect has an electronic configuration of $a_1^{\uparrow\downarrow}\tilde{a}_1^{\uparrow}e^{\uparrow\uparrow}$, where the two non-degenerate defect states ($a_1$ and $\tilde{a}_1$), and the doubly-degenerate states ($e$) are the four symmetry-adapted molecular orbitals formed from the dangling sigma-bonds of the four carbons surrounding the defect [40]. The "tilde" on $\tilde{a}_1$ is used to distinguish it from the other non-degenerate defect state, $a_1$. These defect states are labelled by their irreducible representations for a defective structure with $C_{3v}$-symmetry. It should be pointed out that only $N_C$ on the $C_3$-axis of $V_{Si}$ leaves the symmetry of $V_{Si}$ unchanged, while the other sites change the symmetry to $C_1$.

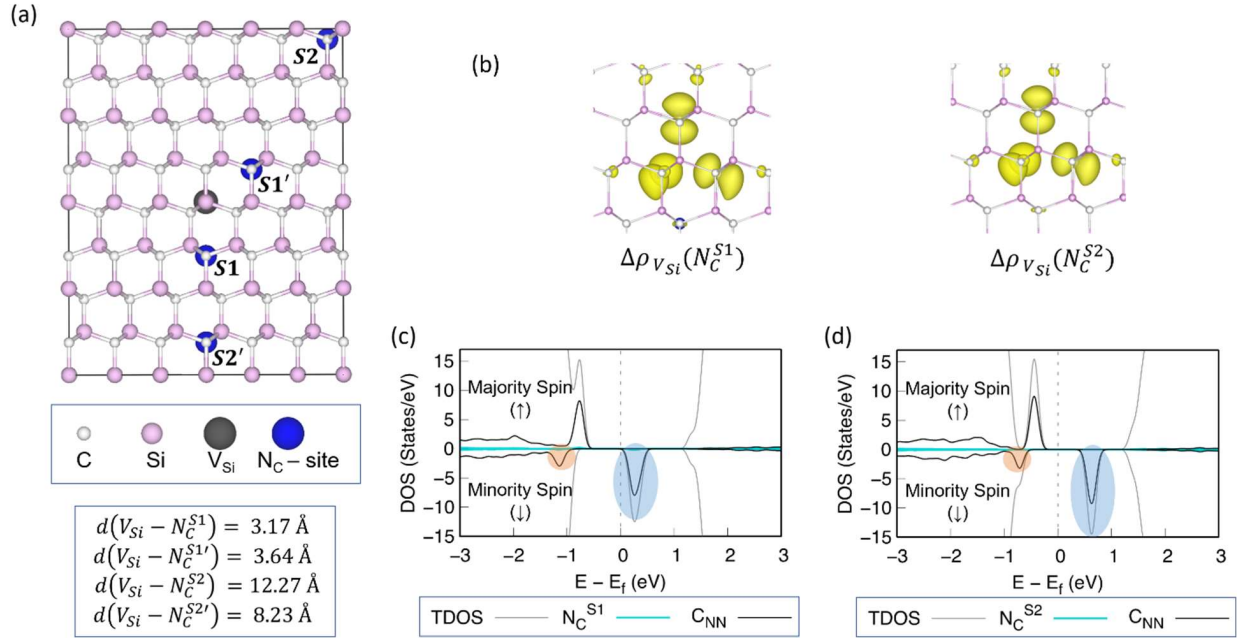

Figure 9. Nitrogen-doping 4H-SiC to stabilize the negative charge state of the silicon monovacancies: (a) Two of the $N_C$ sites investigated are closer to $V_{Si}$ ($S1, S1'$) and two sites are farther away from the monovacancy ($S2, S2'$). Pink spheres are silicon atoms and white spheres are carbon atoms. Silicon monovacancy and nitrogen substitutional sites are marked with large grey and blue spheres, respectively. (b) Spin density ($\Delta\rho_{V_{Si}} = \rho^\uparrow - \rho^\downarrow$) isosurface plots show that most of the spin (3/2) is contributed by the four nearest-neighboring carbon-atoms around $V_{Si}^{-1}$. (c) Density of states (DOS) plot for $V_{Si}$ in 4H-SiC, which is doped by $N_C$ at the S1 site. Along with the total DOS (TDOS), we have also shown the DOS projected onto $N_C$ and the four carbons surrounding $V_{Si}$, labelled as $C_{NN}$. The defect states in the optically-active, minority-spin channel (taken to be the spin-down channel) are highlighted in orange (filled-state) and blue (empty-states). The filled, spin-down defect state (orange highlight) is resonant with the valence band. The DOS plot for $N_C$ at the $S1'$-site is qualitatively similar and not shown. (d) DOS plot for $V_{Si}$ with $N_C$ at the S2 site. The filled, spin-down defect state (orange highlight) in this case lies above the valence band edge. The DOS plot for $N_C$ at the $S2'$-site is qualitatively similar and not shown.

The spin density ($\Delta\rho_{V_{Si}} = \rho^\uparrow - \rho^\downarrow$) is plotted in Fig. 9(b) for the nitrogen-dopant at the $S1$- and $S2$- sites. Most of the spin associated with $V_{Si}$ is localized on the four nearest-neighboring carbons ($C_{NN}$) surrounding $V_{Si}$. This can also be seen in the density of states (DOS) plots shown in Figs 8(c) and (d). Qualitatively, the behavior of $V_{Si}$ with $N_C$ at the $S1'$ ($S2'$) site is the same as when the nitrogen dopant is at the $S1$ ($S2$)-site. We have therefore shown the results for the $S1$ and $S2$-sites only. Figure 9(c) shows that the filled, optically active, minority-spin defect state (highlighted in orange) becomes resonant with the valence band when the dopant is closer to the vacancy. Hence, for monovacancies with nearby nitrogen dopants, the bulk-like states at the valence-band edge will become involved in the excitation process, possibly resulting in deterioration of the optical signal from such a $V_{Si}$. This is consistent with the experimental results in Fig. 3(a-c), in which higher N-doped samples show a larger decrease in PL with increasing annealing temperature beyond 600 °C. Higher temperatures can result in an increased migration of the $V_{Si}$ and $N_C$ towards each other to form more energetically-favorable configurations, either resulting in the filled defect state becoming resonant with the valence band and/or forming an NV-center. In either case, the PL associated with the V2-center will be reduced. In the case of $V_{Si}$ with $N_C$ at the farther $S2$-site [see Fig. 9(d)], the optically-active defect states remain in the bandgap, and the defect retains its properties.

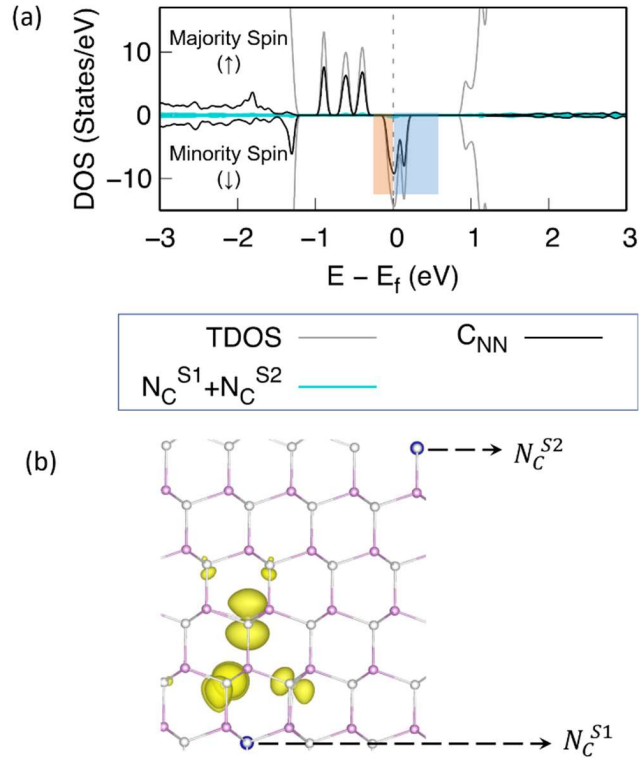

Figure 10. $V_{Si}$ stabilized in the doubly-charged state in the presence of two nitrogen-dopants: (a) Density of states (DOS) plot for $V_{Si}^{-2}$ in 4H-SiC, with nitrogen-substituents at the S1 and S2 sites. Along with the total DOS (TDOS), we have also shown the DOS projected onto the nitrogen substitutionals and the four carbons surrounding $V_{Si}$ (labelled as $C_{NN}$). The defect states that will be involved in the lowest-energy, spin preserving excitations in the minority-spin channel (taken to be the spin-down channel) are highlighted in orange (filled state) and blue (empty state). (b) Spin density ($\Delta\rho_{V_{Si}}$) plot shows that most of the spin is contributed by the four nearest-neighboring carbon-atoms around $V_{Si}^{-2}$, but with spin-distribution that is distinct from that for $V_{Si}^{-1}$.

In order to further understand why the PL from $V_{Si}$ decreases for very high nitrogen-doping levels, we also studied the properties of $V_{Si}$ in the presence of two nitrogen dopants (at $S1$ and $S2$). With the increased level of dopants, we find that $V_{Si}$ is doubly-charged and has a total of six electrons in the dangling bonds. As a consequence, $V_{Si}^{-2}$ is a spin-1 defect. Figure 10 summarizes the electronic [see the DOS plot in Fig. 10(a)] and spin properties of $V_{Si}^{-2}$, with the spin density plot in Fig 10(b) showing that most of the spin-1 is contributed by the four carbons surrounding the defect. Due to the different electronic, spin and optical properties of $V_{Si}^{-2}$, this charge state would not contribute to the PL associated with the V2-center, providing an explanation for the reduced PL observed for the highly N-doped SiC.

## V. CONCLUSION

Nitrogen doping of SiC plays a significant role in the PL, ODMR, and coherence properties of $V_{Si}$ in SiC. Doping at a level commensurate with the irradiation dose can optimize the PL signal, which is consistent with theoretical calculations of the charge state of $V_{Si}$ with nearby $N_C$. Doping too high results in multiple $N_C$ near $V_{Si}$, giving rise to the higher charge states that do not luminesce in the wavelengths detected. Somewhat surprisingly, moderate annealing of SiC at higher doping levels results in a significant improvement in contrast over lower doping, without much reduction in PL. This improvement could be due to reduced lattice damage with annealing, resulting in more unperturbed $V_{Si}$ that have spin transitions and ZPL at the expected values. The role that nitrogen plays in this process is unknown. One might also consider changes in the intersystem crossing as $N_C$ approaches $V_{Si}$ during annealing, which could affect contrast. Further study is needed to determine the mechanism for the improvement. The presence of significant concentrations of electron spins and nuclear spins from nitrogen comes at the cost of spin coherence times. Despite this, there is still an expected improvement in cw ODMR magnetic field sensitivity by a factor of ~1.7 at the highest doping level. These results improve our understanding of this defect, give improvements in magnetic field sensitivity, and perhaps, with further research, lead to approaches for further improvements for defect-based sensing.